\documentclass{iopart}

\usepackage[hypertex]{hyperref}\usepackage{srctex}
\usepackage{graphicx}
\usepackage{iopams}
\usepackage{amssymb}
\newcommand{\RS}{{\scriptscriptstyle{\rm RS}}}
\newcommand{\RSB}{{\scriptscriptstyle{\rm 1RSB}}}
\newcommand{\RSBt}{{\scriptscriptstyle{\rm 2RSB}}}
\newcommand{\RSBn}{{\scriptscriptstyle{\rm nRSB}}}
\newcommand{\mRSB}{\mathit m_{\scriptscriptstyle{\rm1 RSB}}}
\newcommand{\wRSB}{\mathit w_{\scriptscriptstyle{\rm1 RSB}}}
\newcommand{\wRSBpo}{\mathit w^{(p-1)}_{\scriptscriptstyle{\rm1 RSB}}}
\newcommand{\wRSBp}{\mathit w^p_{\scriptscriptstyle{\rm1 RSB}}}
\newcommand{\xRSB}{\mathit x_{\scriptscriptstyle{\rm1 RSB}}}
\newcommand{\mmRSB}{\mathit m^2_{\scriptscriptstyle{\rm1 RSB}}}
\newcommand{\vRSB}{\mathit v_{\scriptscriptstyle{\rm1 RSB}}}
\newcommand{\vvRSB}{\mathit v^2_{\scriptscriptstyle{\rm1 RSB}}}
\newcommand{\vvvRSB}{\mathit v^3_{\scriptscriptstyle{\rm1 RSB}}}
\newcommand{\rRSB}{\mathit r_{\scriptscriptstyle{\rm1 RSB}}}
\newcommand{\rRSBp}{\mathit r^{(p-1)}_{\scriptscriptstyle{\rm1 RSB}}}
\newcommand{\qRS}{\mathit q_{\scriptscriptstyle{\,\rm RS}}}
\newcommand{\qRSpo}{\mathit q^{(p-1)}_{\scriptscriptstyle{\,\rm RS}}}
\begin{document}

\title{Continuous and discontinuous transitions in generalized p-spin glass models}

\author{E.~E.~Tareyeva}
\address{Institute for High Pressure Physics, Russian Academy of Sciences, Troitsk 142190,
Moscow, Russia}
\author{T.~I.~Schelkacheva}
\address{Institute for High Pressure Physics, Russian Academy of Sciences, Troitsk 142190,
Moscow, Russia}
\author{N.~M.~Chtchelkatchev}
\address{Institute for High Pressure Physics, Russian Academy of Sciences, Troitsk 142190,
Moscow, Russia}
\address{Department of Theoretical Physics, Moscow Institute of Physics and Technology, 141700 Moscow, Russia}
\address{Department of Physics and Astronomy, California State University Northridge, Northridge, CA 91330, USA}
\begin{abstract}
We  investigate the generalized $p$-spin models that contain arbitrary
diagonal operators $\hat U$ with no reflection symmetry. We derive general equations that give an opportunity to uncover the behavior of the system near the glass transition at different (continuous) $p$. The quadrupole glass with ${\bf J}=1$ is considered as an illustrating example. It is shown that the crossover from continuous to discontinuous glass transition to one-step replica breaking solution takes place at $p=3.3$ for this model. For $p <2+\Delta p$, where  $ \Delta p= 0.5$ is a finite value, stable 1RSB-solution disappears. This behaviour is strongly different from that of the $p$-spin Ising glass model.
\end{abstract}

\maketitle

\section{Introduction\label{Sec:Intro}}

The theory of spin glasses has been introduced  as an attempt to describe unordered equilibrium freezing of spins in actual dilute magnetic systems with disorder and frustration. This problem was soon partially solved at the mean-field level~\cite{edwards1975JPF,sk1975}: using replica trick to average over disorder  the replica-symmetric (RS) solution was obtained. However, it was soon shown that an adequate description of the low-temperature phase requires a breaking of  the replica symmetry. Already in the paper of Sherrington and Kirkpatrick~\cite{sk1975} (SK) it was shown  that the replica symmetric ansatz is not the correct one. It leads to a negative zero temperature entropy. In the subsequent paper of de Almeida and Thouless of 1978~\cite{Almeida1978} it was shown that the replica symmetric unsatz gives an unstable solution in all the low temperature phase, and hence calling for a replica symmetry breaking. Different approaches to replica symmetry breaking (RSB) were considered (see, e.g. \cite{blandin1978JPhysique,bray1978PRL,Parisi1979}). Two group breaking was proposed by Bray and Moore~\cite{bray1978PRL}. Parisi introduced the method of replica symmetry breaking  step by step with the limit --- full RSB (FRSB) when glass order parameter becomes a continuous non-decreasing function $q(x)$ of a parameter $0\leq x\leq 1$. It provides the hierarchical distribution of pure states overlaps probability $P(q)$ through $P(q) = dx/dq$~\cite{Parisi1979,mezard1987spin}.

Now it is largely believed that similar approach  occurs in general glass-models. However the details of the particular RSB-scheme in different spin-like-glasses and its dependence on the properties of models are far from being understood and remain to be in the focus of intense investigations~\cite{gillin2001multispin,GillinSherringtonJphys2001Comment,Parisi2003,Montanari2003,Crisanti2005,Crisanti2006,Crisanti2007,zamponi2010,crisanti2010sherrington,JanisPRB2011,Janis2013}.

The $p$-spin  spin-glass model of randomly interacting $p$ Ising spins was introduced as a natural generalization of the SK model~\cite{Derrida1980,Derrida1981,Gross1984}. It was shown that in the limit $p\to\infty$ the first step of RSB (1RSB) gives a solution which  remains stable down to zero temperature. In Ref.~\cite{Gardner1985} the detailed investigation of the model was done for $p>2$. It was shown in particular for $p=2+\epsilon $ with small $\epsilon $ that 1RSB solution appears at some $T_c$ with a jump in the order parameter and remains stable in some interval near $T_c$. Both these values, the jump at $T_c$ and the range of 1RSB solution stability, tend to zero as $p\to2$.

Similar behavior as in the $p\to\infty$ model of interacting hard Ising spins was observed in the spherical $p$-spin models with $p>2$~\cite{crisanti1992sphericalp}. These models, as well as soft Potts models, were supposed to be the prototypes of structural glasses~\cite{Kirkpatrick1987,KirkpatrickThirumalai1987,KirkpatrickWolynes1987}. It is worthwhile noting that in Potts glass the jump of the order parameter at the transition to 1RSB state exists beginning at $p=5$~\cite{Gribova2010} while the 1RSB solution remains stable in a pronounced interval already for $p=3$~\cite{GribovaPRE2003}. As to ``spherical'' Potts model with $p=3$ its ground state is replica symmetric~\cite{gribovaPLA2006} as in spherical SK model~\cite{Kosterlitz1976}.

\begin{figure}[b]
  \centering
  \includegraphics[width=0.6\textwidth]{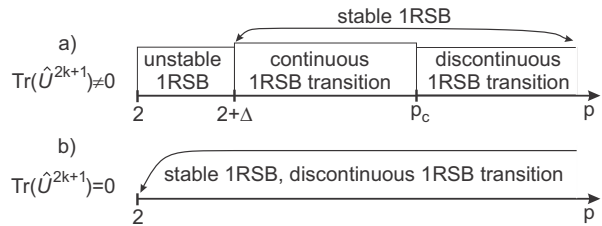}\\
  \centering\caption{The sketch of the ``phase diagram'' of the generalized $p$-operator model, (a) without reflection symmetry (our main result) and (b) with reflection symmetry (this is, e.g., $p$-spin glass model of randomly interacting $p$ Ising spins). Here $k$ is integer while $\Delta$ and $p_c$ are model-dependent constants that we find analytically and numerically for a $p$-quadrupole model.  }\label{fig0}
\end{figure}

It should be noted  that for a long time the discontinuity of the order parameter at the transition and the stability of 1RSB solution were associated with the absence of time-reversal (reflection) symmetry. This problem is still relevant today~\cite{ParisiPRE1999,BanosPNS2012}. Usual paradigm is that the absence of reflection symmetry should be incorporated into the structure of the Hamiltonian. However, it can sometimes be caused by the characteristics of the interacting operators themselves. This was so in our recent investigations, see, e.g., Refs.~\cite{SchelkachevaPRB2010,SchelkachevaJPhysA2011}. We can consider a generalization of the spin model of Ising spins where arbitrary diagonal operators $\hat{U}$ stand instead of Ising spins. In such a way a number of real physical systems can be described, see, e.g. the reviews~\cite{SchelkachevaPRB2010,TareyevaTMF2009}. The operators have different physical origin depending on the problem under study. For example, Ising spin should be replaced with the molecule multipole moment if freezing of the orientational order is the target of the investigation~\cite{SchelkachevaPRB2010,TareyevaTMF2009,Walasek1995,SchelkachevaPRE2009}.

\begin{figure}
  \centering
  \includegraphics[width=0.8\textwidth]{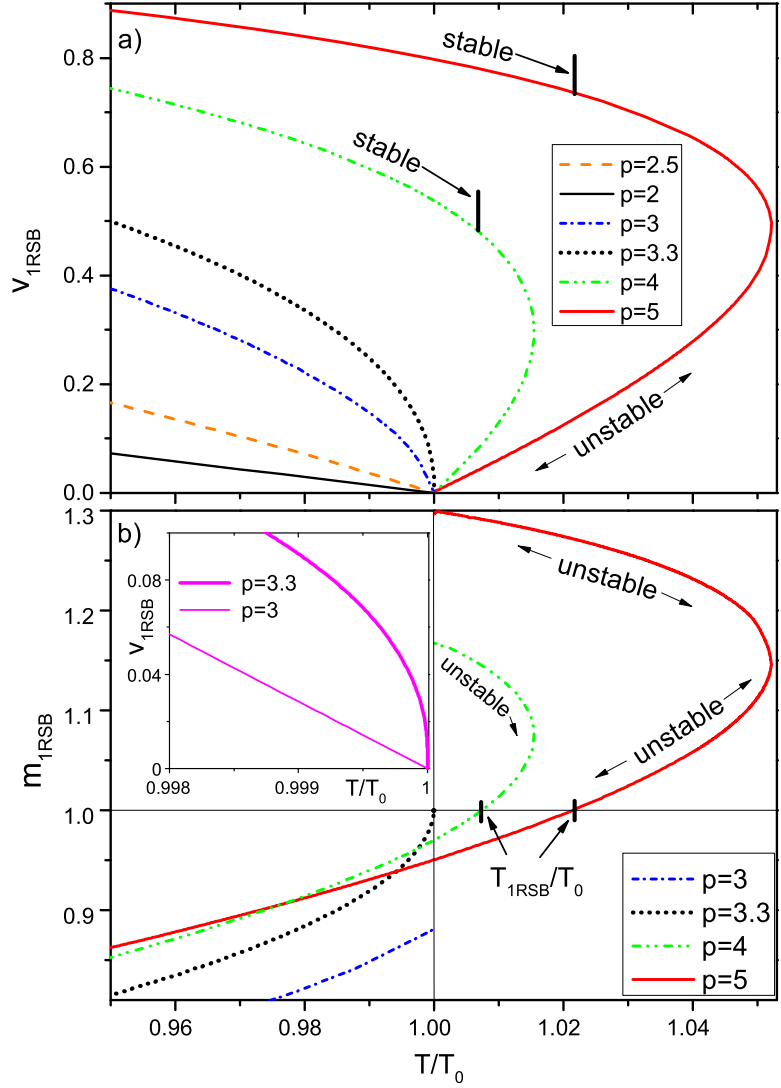}\\
  \caption{(Color online) Graph (a) and inset in (b) show temperature dependence of the glass 1RSB order parameter $\vRSB$ while (b) corresponds to 1RSB order parameter $\mRSB$ for the $p$-quadrupole model. In the case in hand, $\hat{U}= {3J_{z}^{2}}-2$ is the quadrupole moment of the molecule with the angular momentum operator $J_z$. Here we focus on $J=1$, so $J_{z} =\{0,\pm1\}$. When $p = 3.3$ the transition from continuous to discontinuous scenario takes place, $\mRSB = 1$ at the branch point $T_{0}$ and $\vRSB\backsim \sqrt{T_{0}- T}$ near $T_{0}$.}\label{fig2}
\end{figure}
Recently the generalized $p$-spin model with $p=3$ has been considered with a kind of quadrupole operators instead of Ising spins~\cite{SchelkachevaPRE2009}. This model describes, in agreement with experiments~\cite{GoncharovPRB1996}, high pressure orientational glass phase in solid molecular ortho-$D_{2}$ and para-$H_{2}$ where the interactions of more than two particles play an important role.

\textit{Our results.} In this paper we investigate the generalized $p$-spin models that contain arbitrary diagonal operators $\hat U$ instead of Ising spins and focus on the case of $\hat U$ that have no reflection symmetry, that is $ \Tr{\hat{U}}^{(2k+1)} \neq 0$ for all integer $k>0$. Now the glass freezing scenario is absolutely different from the Ising $p$-spin case. There is no temperature interval, in particular, where RS-solution coincides with the ``para-state'' of pure system.

We obtain general equations for order parameters near the glass transition for arbitrary real $p>2$ using expansions of the effective free energy and the bifurcation theory~\cite{VainbergTrenogin}. That allowed us to uncover the behavior of the generalized $p$-spin models depending on the continuous parameter $p$ and the symmetry properties of the operator $\hat{U}$. This is one of main results of our paper and we represent it qualitatively in figure~\ref{fig0}a. We have found analytically and numerically the critical value of $p$  where a crossover from continuous to discontinuous transition to 1RSB solution takes place.
In the case $ \Tr{\hat{U}}^{(2k+1)} \neq 0$, the problem has the property of continuity in the parameter $p$. The results we have found differ from the corresponding results for the case $\Tr{\hat{U}}^{(2k+1)}=0$~\cite{SchelkachevaJPhysA2011} when  point $p=2$ is a distinguished point, see figure~\ref{fig0}b.

\begin{figure}[t]
  \centering
  \includegraphics[width=0.8\textwidth]{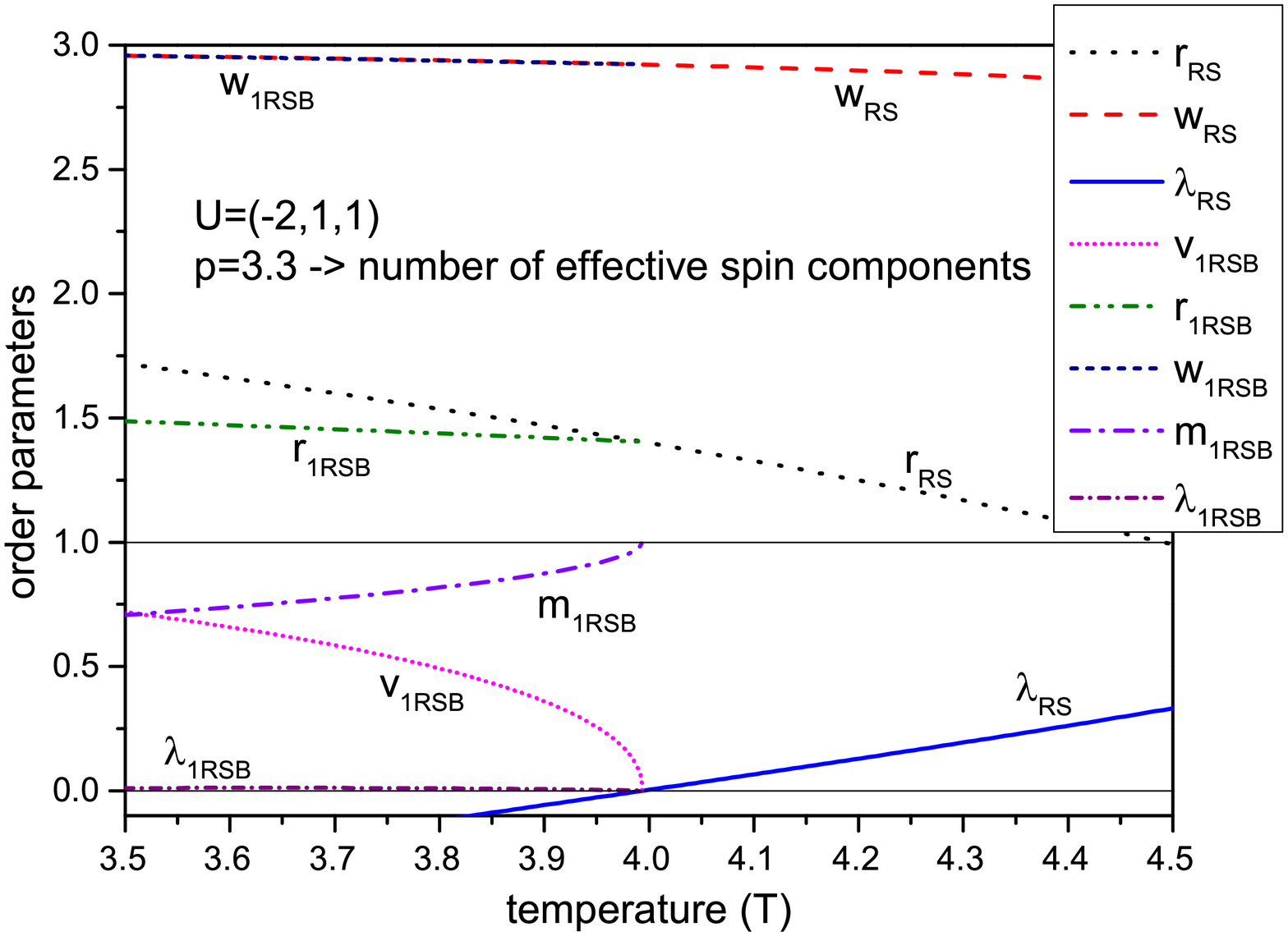}\\
  \caption{(Color online) Temperature dependence of the order parameters, $\lambda_{\rm (\RS) repl}$ and $\lambda_{\rm (\RSB) repl}$ of the  $p$-quadrupole  model for the boundary value $p=3.3$ where the crossover from continuous to discontinuous 1RSB transition takes place. We should emphasize that 1RSB order parameter $\vRSB$ approaches zero nonlinearly.
}\label{fig1}
\end{figure}

As an illustrative example we consider the quadrupole glass with ${\bf J}=1$.  We investigate temperature dependence of the order parameters, see figures~\ref{fig2}-\ref{fig1}, scan the range of 1RSB stability depending on $p$ for this model, see figure~\ref{fig3}. Stable 1RSB-solution appears for $p>2.5$. We show that the crossover from continuous to discontinuous transition occurs at $p=3.3$, see figure~\ref{fig2}.

\textit{Structure of the paper.} In Sec.~\ref{Sec:Maineq} we write down the standard starting equations for the generalized p-spin model and discuss them in Sec.~\ref{sec:discussion}. In Sec.~\ref{sec:Stability} we investigate the stability and examine the continuity depending on the parameter $p$. We write down general expansions of the effective free energy near the transition (as was done, e.g., in Ref.~\cite{SchelkachevaPRB2010,SchelkachevaJPhysA2011}) and finally arrive at Eqs.~(\ref{602prs})-(\ref{B73RS}). These expressions lead us to the ``phase diagram'' of the generalized $p$-operator model shown in figure~\ref{fig0} that schematically shows main results of our paper. In Sec.~\ref{sec:quadrupole} we consider as an illustrative example the $p$-quadrupole glass. In Sec.~\ref{Sec:Conc} we write the conclusions.

\section{ Generalized p-spin model \label{Sec:glass}}

\subsection{Main equations \label{Sec:Maineq}}
The Hamiltonian of the $p$-operator model in general looks like:
\begin{equation}
H=-\sum_{{i_{1}}\leq{i_{2}}...\leq{i_{p}}}J_{i_{1}...i_{p}}
\hat{U}_{i_{1}}\hat{U}_{i_{2}}...\hat{U}_{i_{p}}, \label{one}
\end{equation}
where $\hat{U}$ now is arbitrary diagonal operator with $\Tr\hat{U}=0 $, $N$ is the number of sites on the lattice, $i=1,2,...N$, and $p$ gives the
number of interacting particles. The coupling strengths are independent random variables with Gaussian distribution
\begin{equation}
P(J_{i_{1}...i_{p}})=\frac{\sqrt{N^{p-1}}}{\sqrt{p!\pi}
J}\exp\left[-\frac{(J_{i_{1}...i_{p}})^{2}N^{p-1}}{ p!J^{2}}\right]. \label{two}
\end{equation}

Using in the standard way the replica trick, see, e.g.~\cite{mezard1987spin}, we write in general the free energy averaged over
disorder~\cite{SchelkachevaJPhysA2011,SchelkachevaPRE2009}:
\begin{eqnarray}
\nonumber\fl\langle F\rangle_J/NkT=\lim_{n \rightarrow 0}\frac{1}{n}\max\left \{(p-1) \frac{t^2}{4}\sum_{\alpha}
(w^{\alpha})^{p} + \right.
\\\left.
(p-1)\frac{t^2}{2}\sum_{\alpha>\beta} (q^{\alpha\beta})^{p}-
\ln\Tr_{\{U^{\alpha}\}}\exp \hat{\theta}\right\}.\label{free}
\end{eqnarray}
where
\begin{equation}
\hat{\theta}=p\frac{t^2}{2}
\sum_{\alpha>\beta}(q^{\alpha\beta})^{(p-1)}\hat{U}^{\alpha}\hat{U}^{\beta}+p\frac{t^2}{4}
 \sum_{\alpha}{(w^{\alpha})}^{(p-1)}(\hat{U}^{\alpha})^2. \label{six}
 \end{equation}
Here $t={J}/kT$.

The saddle point conditions give the glass order parameter
\begin{equation}\label{qdef}
q^{\alpha\beta}={\Tr\left[\hat{U}^{\alpha}\hat{U}^{\beta} \exp\left(\hat{\theta}\right)\right]}/{\Tr\left[\exp\left(\hat{\theta}\right)\right]},
\end{equation}
and the auxiliary order parameter
\begin{equation}\label{wdef}
 w^{\alpha}={\Tr\left[(\hat{U}^{\alpha})^2 \exp\left(\hat{\theta}\right)\right]}/{\Tr\left[\exp\left(\hat{\theta}\right)\right]}.
\end{equation}
We also introduce by analogy the regular order parameter
\begin{equation}\label{xdef}
 x^{\alpha}={\Tr\left[(\hat{U}^{\alpha})
\exp\left(\hat{\theta}\right)\right]}/{\Tr\left[\exp\left(\hat{\theta}\right)\right]}.
\end{equation}

 Using the standard procedure~\cite{mezard1987spin} we perform
the first stage of the replica symmetry breaking  ($n$ replicas are
divided into  $n/\mRSB$ groups with  $\mRSB$ replicas in each) and obtain
the expression for the free energy. Order parameters are denoted by
$q^{\alpha \beta }= \rRSB$ if $\alpha $ and $\beta $ are from different
groups and $q^{\alpha \beta }= \rRSB+\vRSB$ if $\alpha $ and $\beta $
belong to the same group. So
\begin{eqnarray}
\nonumber\fl F_{\rm \RSB}=-NkT\left\{\mRSB t^2(p-1)\frac{\rRSB^p}{4}+(1-\mRSB)(p-1)
t^2\frac{(\rRSB+\vRSB)^p}{4}- \right.
\\
\left. t^2(p-1)\frac{\wRSBp}{4}+\frac{1}{\mRSB}\int dz^G\ln
\int ds^G
\left[\Tr\exp\left(\hat{\theta}_{\RSB}\right)\right]^{\mRSB}\right\}.
\label{frs}
\end{eqnarray}
Here
\begin{eqnarray}
\nonumber\fl \hat{\theta}_{\rm \RSB}=\left.z\,t\sqrt{\frac{p\,\rRSBp}{2}}\,\hat{U}+st\sqrt{\frac{p\,[{(\rRSB+\vRSB)}^{(p-1)}-\rRSBp]}{2}}\,\hat{U}+
\right.
\\
\left.
t^2\frac{p\,[\wRSBpo-{(\rRSB+\vRSB)}^{(p-1)}]}{4}\hat{U}^2,\right.
\end{eqnarray}
and
$\int dz^G = \int_{-\infty}^{\infty} \frac{dz}{\sqrt{2\pi}}\exp\left(-\frac{z^2}{2}\right)$.
 The extremum conditions for $F_{\rm \RSB}$ yield equations for the glass
order parameters $\rRSB$ and $\vRSB$, the additional order parameter
$\wRSB$, the regular order parameter $\xRSB$ and the parameter $\mRSB$:
\begin{eqnarray}\label{18qrs}
\fl \rRSB=\int dz^G\left\{ \frac{\int ds^G{\left[\Tr\exp\hat{\theta}_{\rm \RSB}\right]}^{(\mRSB-1)}\left[\Tr\hat{U} \exp\hat{\theta}_{\rm \RSB}\right]}
{\int ds^G{\left[\Tr\exp\hat{\theta}_{\rm \RSB}\right]}^{\mRSB}}\right\}^{2},
\\\nonumber\fl
\vRSB=
\int dz^G \frac{\int ds^G{\left[\Tr\exp\hat{\theta}_{\rm \RSB}\right]}^{(\mRSB-2)}{\left[\Tr{\hat{U} }\exp\hat{\theta}_{\rm \RSB}\right]}^{2}}
{\int ds^G{\left[\Tr\exp\hat{\theta}_{\rm \RSB}\right]}^{\mRSB}}   -
\\
\int dz^G\left\{ \frac{\int ds^G{\left[\Tr\exp\hat{\theta}_{\rm \RSB}\right]}^{(\mRSB-1)}\left[\Tr\hat{U} \exp\hat{\theta}_{\rm \RSB}\right]}{\int ds^G{\left[\Tr\exp\hat{\theta}_{\rm \RSB}\right]}^{\mRSB}}\right\}^{2},\label{1vrs}
\\\label{1prs}
\fl \wRSB=
\int dz^G \frac{\int ds^G{\left[\Tr\exp\hat{\theta}_{\rm \RSB}\right]}^{(\mRSB-1)}\left[\Tr{\hat{U} }^{2}\exp\hat{\theta}_{\rm \RSB}\right]}
{\int ds^G{\left[\Tr\exp\hat{\theta}_{\rm \RSB}\right]}^{\mRSB}},
\end{eqnarray}
\begin{eqnarray}\nonumber\fl
\mRSB\frac{t^{2}}{4}(p-1)\left[{(\rRSB+\vRSB)}^{p}-{(\rRSB)}^{p}\right]=
\\
- \frac{1}{\mRSB}\int dz^G\ln\int ds^G \left[\Tr\exp\hat{\theta}_{\rm \RSB}\right]^{\mRSB}+
\\\nonumber
\int dz^G \frac{\int ds^G{\left[\Tr\exp\hat{\theta}_{\rm \RSB}\right]}^{\mRSB}\ln\left[\Tr\exp\hat{\theta}_{\rm \RSB}\right]}{\int ds^G{\left[\Tr\exp\hat{\theta}_{\rm \RSB}\right]}^{\mRSB}}.\label{31mrs}
\end{eqnarray}

The expressions for the RS-approximation can be found technically  from Eqs.~(\ref{18qrs})-(\ref{1prs}) setting $\vRSB = 0$. Then for glass order parameter $q_{\rm \RS}$ we have:
\begin{equation}\label{0qrs}
\qRS=\int dz^G\left\{ \frac{\Tr\left[\hat{U} \exp\left(\hat{\theta}_{\rm \RS}\right)\right]}
{\Tr\left[\exp\left(\hat{\theta}_{\rm \RS}\right)\right]}\right\}^{2},
\end{equation}
where
\begin{equation}\label{1qrs}
\hat\theta_{\rm \RS}=z\,t\sqrt{\frac{p\,\qRSpo}{2}}\,\hat{U}+t^2\frac{p\,[w_{\rm \RS}^{(p-1)}-
\qRSpo]}{4}\hat{U}^2.
\end{equation}

\subsection{Discussion \label{sec:discussion}}
We should emphasize that  the characteristic properties of the systems in hand develop themselves already in the replica symmetric (RS) approximation. If $\hat U$ has \textit{no reflection symmetry} then the nonlinear integral equation for the RS-glass order parameter~(\ref{0qrs})-(\ref{1qrs}) simply has no trivial solutions at any temperature because the integrand is nonsymmetric due to the cubic terms in the free-energy expansion~\cite{SchelkachevaPRB2010,TareyevaTMF2009}. Both physical RS order parameters, glass ~(\ref{qdef}) and regular~(\ref{xdef}),  increase smoothly as the temperature decreases from the high temperature nonzero values (see also below Section 2.4). There is no temperature interval where RS-solution coincides with the ``para-state'' of pure system.  At the point where the RS-solution becomes unstable (the bifurcation point  $T_{0}$ of the Eq.~(\ref{1vrs}) for the glass order parameter) 1RSB branch appears. 1RSB glass parameters $\vRSB$ and $\rRSB$ appear continuously at the bifurcation point  $T_{0}$.

We find 1RSB order parameter $\mRSB$ analytically at the branching point $T_{0}$ and express the result through RS-order parameters. If $\mRSB\leq1$ at the branching point then 1RSB solution has physical meaning (if stable) near $T_{0}$, see, e.g., figure~\ref{fig2}. The condition $\mRSB>1$ at $T_{0}$ implies the unphysical region and so then $T_{0}$ is not the actual point of the transition. There 1RSB glass parameters show the recurrent behavior with many-valued dependence on temperature. As the result we have a jump of physical glass order parameters at $T_{\rm 1RSB}$ where $\mRSB=1$, see~figure~\ref{fig2}. In some sense 1RSB solution behaves here like in ordinary $p$-spin model  but in the external field~\cite{Oliveira1999replica}. This is due to the fact that the subalgebra of commuting diagonal matrices to which belongs our $\hat U$ necessarily has the dimension greater than two, so that  ${\hat U}^2$ can contain $\hat U$ (or some other operator of the subalgebra).

In contrast, \textit{reflection symmetry} of the operators $\hat{U}$ should lead to vanishing of a number of terms in the free energy, so that the replica symmetric (RS) solution for the order parameters is zero at high temperature. As the result the behavior of 1RSB solution for the order parameters is like in the ordinary $p$-spin model of Ising spins. In this case for $p=2$ we have FRSB appearing just in the transition point~\cite{schelkachevaPLA2006} and ultrametricity of the space of states~\cite{VasinTMF2013}.

\subsection{Stability of 1RSB solution\label{sec:Stability}}

\subsubsection{Replicon mode}
The stability of RS and nRSB solutions can be tested by the investigation of the gaussian fluctuation contribution to the free energy near this solution, see, e.g.~\cite{mezard1987spin}. The saddle point solution is stable if all the eigen modes of the fluctuation propagator are positive. The most important mode is the so-called replicon mode~\cite{Almeida1978,TareyevaTMF2009} since its sign is usually very sensitive to the replica symmetry breaking and to the temperature. It is just $\lambda_{\rm (\RSBn) repl}$ that enter the free energy with $v^{2}_{\RSBn}$, where $v_{\RSBn}$ define the novel intragroup difference of the order parameters~\cite{SchelkachevaJPhysA2011}. For example, the replica symmetric solution is stable unless the corresponding replicon mode energy ${\lambda_{\rm (\RS) repl}}>0$. The RS-solution can break at the temperature $T_{0}$ determined by the equation $\lambda_{\rm (\RS) repl}=0$, where
\begin{equation}\label{lambdaRS}
\fl\lambda_{\rm (\RS) repl}= 1 - t^{2} \frac{p(p-1)q_{\rm \RS}^{(p-2)}}{2}
\int dz^G \left\{\frac{\Tr\left(\hat{U}^2
e^{\hat{\theta}_{\rm \RS}}\right)} {\Tr e^{\hat{\theta}_{\rm \RS}}}-
\left[\frac{\Tr\hat{U} e^{\hat{\theta}_{\rm RS}}}
{\Tr e^{\hat{\theta}_{\rm RS}}}\right]^2\right\}^2.
\end{equation}
The equation $\lambda_{\rm (\RS) repl}=0$ can be obtained as the branching condition for (\ref{1vrs}), i.e., as the condition that a small solution with 1RSB can appear.

From Eq.~(\ref{lambdaRS}) follows, in particular, that the point $p=2$ is a singular point in the case $\Tr{\hat{U}}^{(2k+1)}=0$ (since $q_{\RS}=0$), see figure~\ref{fig0}.  Let us recall that we are investigating
now the case of the absence of reflection symmetry, so that the case
$q_{\RS}=0$ is excluded.

In our case $\Tr{\hat{U}}^{(2k+1)}\neq0$ for $k>0$ and the high-temperature expansion of the equation for the order parameter $q_{\RS}$ does not give trivial solution. Using the condition, $q_{\RS}\neq 0$,  we can find that $T_{0}\neq0$ in contrast with the the reflection symmetry case. Then the solutions with the unbroken symmetry may appear continuously not only for $p=2$.

It is important that  $q_{\RS}=0$ is excluded when we investigate the case without the reflection symmetry of $\hat{U}$. In this case, the problem has the property of continuity over the parameter $p$.

\begin{figure}[t]
  \centering
  \includegraphics[width=0.8\textwidth]{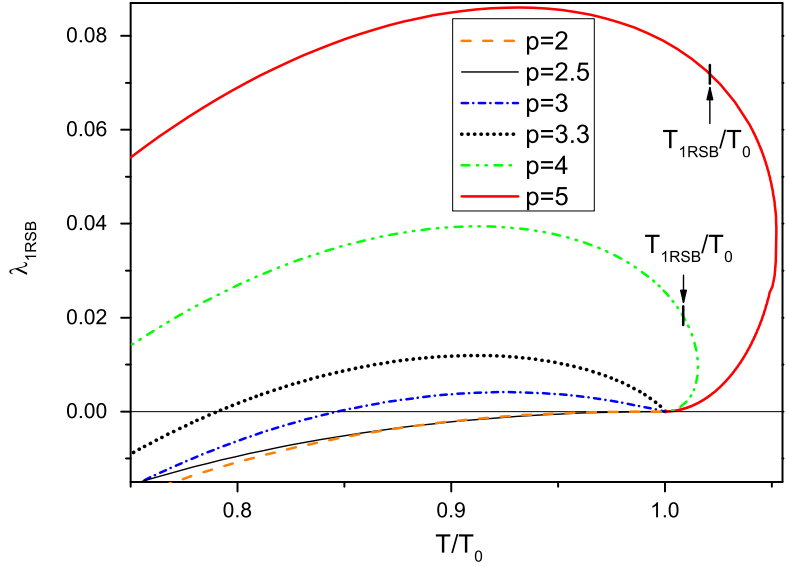}\\
  \caption{(Color online) Temperature dependence of  $\lambda_{\rm (\RS) repl}$ and $\lambda_{\rm (\RSB) repl}$ for $p$-quadrupole glass.  For $p>2.5$ there is a domain where $\lambda_{\rm (\RSB) repl}>0$ and so stable 1RSB-solution appears. At $p = 3.3$ takes place the transition from continuous to discontinuous (jumpwise) scenario of the order parameters evolution with temperature. }\label{fig3}
\end{figure}

If we break the replica symmetry once more then we obtain the corresponding expressions for the free energy and the order parameters. The bifurcation condition $\lambda_{\rm (\RSB) repl}=0$ that determines the temperature $T = T_{2}$ follows from the condition that a nontrivial small solution for the 2RSB intragroup glass order parameter appears as  $v_{ \RSBt}\rightarrow 0$. So,
\begin{eqnarray}
\nonumber\fl\lambda_{\rm (\RSB) repl}= 1 - t^{2} \frac{p(p-1)(\rRSB+\vRSB)^{(p-2)}}{2} \times
\\
\fl\int dz^G \frac{\int ds^G\left[\Tr\exp\left(\hat{\theta}_{\RSB}\right)\right]^{\mRSB}
\left\{\frac{\Tr\left[\hat{U}^2
\exp\left(\hat{\theta}_{\RSB}\right)\right]} {\Tr\left[\exp\left(\hat{\theta}_{\RSB}\right)\right]}-
\left[\frac{\Tr\left[\hat{U} \exp\left(\hat{\theta}_{\RSB}\right)\right]}
{\Tr\left[\exp\left(\hat{\theta}_{\RSB}\right)\right]}\right]^2\right\}^2}{\int ds^G
\left[\Tr\exp\left(\hat{\theta}_{\RSB}\right)\right]^{\mRSB}}\,.\label{lambda}
\end{eqnarray}
Note that Eq.~(\ref{lambda}) depends only on  1RSB-solution. The equation  $\lambda_{\rm (\RSB) repl}=0$ always has the solution for $\vRSB=0$, which determines the point $T_{0}$ and coincides with the solution of Eq.~(\ref{lambdaRS}) $\lambda_{\rm (\RS) repl}=0$, see figure~\ref{fig3}.

In addition to the point $T_{0}$, one more bifurcation point ($T_2$) defined by $\lambda_{\rm (\RSB) repl}=0$ may exist as $\vRSB\neq0$, and the 2RSB solution can appear at this point. A transition to FRSB-state or to a stable 2RSB-state may take place at the point $T_{2}$.

\subsubsection{Free energy expansion}
We expand the free energy (\ref{free})-(\ref{six}) up to the third order to get in general 1RSB solution near the bifurcation  point $T_0$ where it slightly deviates from the RS-solution. We assume that the deviations $\delta q^{\alpha \beta}$ from $q_{\rm \RS}$ and $\rho$ from $w_{\rm \RS}$ are small.  We use the notation $\Delta F$ for the difference between the free energy $F_{\rm (\RSB)}$ and its replica symmetric value  ${F^{\rm (\RSB)}_{0}}$. So,
\begin{eqnarray}\nonumber
\fl\frac{\Delta F}{NkT}=\frac{t^2}{4}\frac{p(p-1)}{2}{q^{(p-2)}_{\RS}} \lambda_{\rm (\RS) repl}\times
\\\nonumber\Biggl\{-\left[r-(\mRSB-1)\vRSB\right]^{2}-\vvRSB\mRSB(1-\mRSB)-
\\\nonumber
\frac{t^{4}}{2}L\left[r-(\mRSB-1)\vRSB\right]^{2}-
t^{6}\biggl\{C\left[r-(\mRSB-1)\vRSB\right]^{3}+
\\\nonumber
D\left[r-(\mRSB-1)\vRSB\right]\vvRSB\mRSB(\mRSB-1)-
\\
B_{3}\vvvRSB\mmRSB(\mRSB-1)+
\\\nonumber
B_{4}\vvvRSB \mRSB(\mRSB-1)(2\mRSB-1)\biggr\}\Biggr\}+\Psi(\rho) +...\label{10frs},
\end{eqnarray}
where $t=t_{0}+\Delta t$, $\rRSB=q_{\rm \RS}+r$, $w_{\rm \RSB}=w_{\rm\RS}+\rho$ and the expressions for $L$, $C$, $D$, $B_3$, $B_4$ and $\Psi $ are some combinations of operators averaged over the RS-solution (the exact expressions can be found here in Appendix and in Appendix of Ref.~\cite{SchelkachevaJPhysA2011}).

Then one can see that $L|_{t=t_{0}}\neq {0}$ for $\Tr{\hat{U}}^{(2k+1)}\neq0$ for $k\neq0$. Using this fact we obtain from the extremum conditions for the free energy (\ref{10frs}) that the branching can take place only if the solution is of the following form
\begin{equation}
r-(\mRSB-1)\vRSB=0+o(\Delta t)^{2}.
\label{4099prs}
\end{equation}
This condition states that there is no linear term for the glass order parameters, see Ref.~\cite{VainbergTrenogin}. There is no other linear term  because $\lambda_{\rm \RS repl}|_{t=t_{0}}=0$ at the bifurcation point. From the extremum condition we get, $\rho\sim [r-(\mRSB-1)\vRSB]$ and $\Psi(\rho)=0+o(\Delta t)^{4}$, see Refs.~\cite{SchelkachevaPRB2010, SchelkachevaJPhysA2011}. Finally, we have:
\begin{eqnarray} \label{30prs}
\fl 2\mRSB(1-\mRSB)Z\Delta
T =3t_{0}^{6}\mRSB(1-\mRSB)\times
\\\nonumber
\left[-B_{4}+\mRSB(-B_{3}+2B_{4})\right]\vRSB\,,
\\\nonumber
\fl (2\mRSB-1)Z\Delta T = t_{0}^{6}\left\{(2\mRSB-1)\left[-B_{4}+\mRSB(-B_{3}+2B_{4})\right]+\right.
\\\left.
\mRSB(\mRSB-1)(-B_{3}+2B_{4})\right\}\vRSB
\,,\label{40prs}
\end{eqnarray}
where
\begin{eqnarray} \label{7850prs}
Z = \frac{t^2}{4}\frac{p(p-1)}{2}\frac{{d[q^{(p-2)}_{\RS}}\lambda_{\rm (\RS) repl}]}{d T }.
\end{eqnarray}
Here $ B_{3}$,  $B_{4}$ and $Z$  are taken at  $T=T_0$. Then we find from (\ref{30prs}) and (\ref{40prs}) at the branch point $T_0$ where 1RSB-solution appears (the cases $\mRSB=0$ and $\mRSB=1 $ should be investigated separately):
\begin{equation}
\mRSB={{B_{4}}/{B_{3}}},\label{50prs}
 \end{equation}
and
\begin{equation}
 \rRSB=q_{\rm \RS}+(\mRSB-1)\vRSB,\qquad
\vRSB=K
\Delta t\label{77prs}
\end{equation}
in the neighborhood of  $T_0$.  The coefficient of proportionality $K$ and $\mRSB$ depends only on RS-solution at $T_{0}$. The exact expression for $K$ is rather lengthy and can be found in Ref.~\cite{SchelkachevaJPhysA2011}. 1RSB solution branches smoothly from RS solution.

From Eqs.~(\ref{30prs})--(\ref{40prs}) and (\ref{50prs}), we obtain
\begin{eqnarray} \label{602prs}
&& \fl 2Z\Delta T=3t_{0}^{6}\mRSB(\mRSB-1)B_{3}\vRSB,
\\\label{B73RS}
&&\fl B_{3}={\left[\frac{p(p-1)}{2}{q^{(p-2)}_{\RS}}\right]}^{3}
\frac{1}{6}\int dz^G \left\{\frac{\Tr\left(\hat{U}^2
e^{\hat{\theta}_{\RS}}\right)} {\Tr e^{\hat{\theta}_{\RS}}}-
\left[\frac{\Tr\hat{U} e^{\hat{\theta}_{\RS}}}
{\Tr e^{\hat{\theta}_{\RS}}}\right]^2\right\}^3\geq0.
\end{eqnarray}
The last inequality follows from the Cauchy--Schwarz inequality $\left(\sum_{n}{A_{n}}^{2} \right)\left(\sum_{n}{B_{n}}^{2}\right)\geq\left(\sum_{n}A_{n}B_{n}\right)^{2} $.

\subsubsection{Discussion}

The qualitative analysis of these solutions near $T_{0}$ we performed using this last set of equations (\ref{7850prs}) - (\ref{602prs}).

From Eq.~(\ref{602prs}) we see that the value of the parameter $\mRSB$ at $T=T_{0}$ determines the behavior of the glass order parameter $\vRSB$. For values ​​of $0<\mRSB<1$ and $\mRSB>1$ there is a linear dependence on $\Delta T$. But the sign of the coefficient of proportionality depends on $\mRSB$. Really, $\lambda_{\rm (\RS) repl}$ is increasing function of temperature, equal to zero at $T_{0}$, and it changes its sign at this point because the sign of $\lambda_{\rm (\RS) repl}$ determines the stability of  RS solution. The order parameter $q_{\rm \RS}$ is positive and slowly varying function of temperature in the vicinity of $T=T_{0}$. So, we have $Z>0$ near $T_{0}$, see~Eq.~(\ref{7850prs}).

In the first case, when $0<\mRSB<1$ the solution goes to the left side of the point $T=T_{0}$ and so the 1RSB solution appears continuously. The resulting solution can be stable only when $\lambda_{\rm (\RSB) repl}>0$. In the second case when $\mRSB>1$ positive solution occurs at $T\geq T_{0}$. The resulting solution is nonphysical solution near $T=T_{0}$ since $\mRSB>1$ has no physical meaning, see figure~\ref{fig2} for illustration. 1RSB solution can only occur abruptly when $\mRSB(T)$ yields the value of  $\mRSB(T)=1$.

If we obtain $\mRSB=1$ at $T=T_{0}$ at a certain value of $p$, thus we find the point where a crossover from continuous to discontinuous transition to 1RSB solution takes place. In this case we see from the equation (\ref{602prs}) that  it is necessary to consider fourth order terms in the expansion of the free energy~(\ref{10frs}). One can show that in this case $v^{2}_{1}\sim \Delta T$, see figure~\ref{fig1} for illustration.

\subsection{Quadrupole glass with $J=1$ \label{sec:quadrupole}}

\subsubsection{Results}

The quadrupolar glass with $J=1$ and  $J_{z} =\{0,\pm1\}$ is the simplest example of the system without reflection symmetry. In this case $\hat{U}= {3J^{2}}_{z}-2$ is the quadrupolar moment of the molecule. It is worth to emphasize that the smooth increasing of glass and regular order parameters from high to low temperature obtained in the frame of the simplest version of such a glass~\cite{lutchinskaiaJPhys1984} was confirmed by the experiment in Ref.~\cite{sullivan1986orientational}. Now we present a calculation based on
Eqs.~(\ref{18qrs})--(\ref{31mrs}) and (\ref{7850prs})--(\ref{602prs}) for the model with varying $p$, see figures~\ref{fig2}-\ref{fig3}.

For  $p=3$, 1RSB solution is stable near $T_{0}$, it branches continuously at the bifurcation point $T_{0} = T_{\rm \RSB}$ and changes smoothly on cooling below $T_{0}$, see figure~\ref{fig2}. We find that $\mRSB = 0.88<1$ at the branch point.~\footnote{We have $Z>0$ near $T_{0}$.} It follows from Eq.~(\ref{602prs}) that $\vRSB\backsim [-(T-T_{0})]$ near the branch point.

For $p = 4$, $p= 5$, \textit{etc}... the solution appears smoothly but in these cases, the condition $\lambda_{\rm (\RS) repl}=0$, does not determine the small physical solution in the neighborhood of the branch point. Namely, $\mRSB> 1$~\footnote{for p=4 we have  $\mRSB=1.17$, for p=5 we have $\mRSB=1.3$} at the branch point and the nonphysical branch of the free energy appears at $T=T_{0}$. So [see Eq.~(\ref{602prs})] the order parameter is linear again:  $\vRSB\backsim (T-T_{0}) $. In fact, the transition from the RS to the 1RSB solution is discontinuous at the point $T_{\rm \RSB} > T_{0}$ determined by the condition $\mRSB = 1$. At this point, $F_{\rm \RS} = F_{\rm \RSB}$. Replica-symmetric solution is stable above $T_{\rm \RSB}$ while $\mRSB<1$ for $T<T_{\rm \RSB}$ and the corresponding physical 1RSB solution corresponds to larger (preferable) free energy than the RS-solution.

At $p = 3.3$ the crossover from continuous to discontinuous scenario takes place. In this case, $\mRSB = 1$ at the branch point. Equation~(\ref{30prs}) becomes an identity. The right-hand sides of Eqs.~(\ref{40prs}) and (\ref{602prs}) become zero. It is therefore necessary to consider the terms of fourth order of glass order parameters in the expansion of the free energy~(\ref{10frs}) near $T_{0}$. Hence now we obtain the nonlinear behaviour of the order parameter near the transition point: $\vRSB\backsim \sqrt{T_{0}- T}$, see figures~\ref{fig2}-\ref{fig1}. It is important that the positivity of $\lambda_{\rm (\RSB) repl}$ is a necessary condition for the solution in these models to be stable with respect to subsequent replica symmetry breaking~\cite{SchelkachevaJPhysA2011}.

The  $p$-quadrupole glass model for $p <2+\Delta p$, where  $ \Delta p= 0.5$ is a finite value, behaves exactly the same way as it takes place in the case of pair interaction. For $p <2.5$ stable 1RSB-solution disappears, see figures~\ref{fig2}-\ref{fig3}. We get $\lambda_{\rm (\RSB) repl}\leq 0$ for  $T<T_{0}$. This behavior is expected due to the continuity reasons.

In our case $\Tr{\hat{U}}^{(2k+1)}\neq0$, the problem has the property of continuity over the parameter $p$. These results differ from the corresponding results for the case $\Tr{\hat{U}}^{(2k+1)}=0$ when the point $p=2$ is a singular point, see Eq.~(\ref{lambdaRS}), since $q_{\RS}=0$.

\subsubsection{Discussion}
Let us note that qualitatively similar behaviour in ceratin aspects shows the Potts model for the 3, 4 and 5 states~\cite{Gribova2010}. The  form of the series for $\Delta F=F_{\RSB}-F_{\RS}$ (\ref{10frs}) over the small deviations $\delta q^{\alpha \beta}$ from $q_{\RS}$ is one and the same for different models. In the case of Potts spin glass model  the reflection symmetry is absent. However, $L=0$,\footnote{We remind that $L$ is defined in Eqs.~(\ref{10frs}) and (\ref{eqL}).} because it is zero RS-solution that bifurcates. As $L|_{t=t_{0}}= {0}$, then the condition $r-(m-1)\vRSB=0$~\footnote{We remind that $\rRSB=q_{\rm \RS}+r$.} is not fulfilled ($\rRSB=r=0$ Ref.~\cite{RitortJPhys1995}). But equations similar to (\ref{30prs}), (\ref{40prs}) and (\ref{50prs}) do exist. Then the crossover, at $p=4$, from continuous to jumpwise behavior with the growing of the number of states can be traced analytically. There also exists a domain of stability where the 1RSB-solution remains stable under further RSBs.~\cite{GribovaPRE2003}

We also remind that it was shown for the $p$-Ising spin-glass model~\cite{Derrida1980,Derrida1981,Gross1984,Gardner1985}  that 1RSB solution appears at some $T_c$ with a jump in the order parameter for $p=2+\epsilon$ with small $\epsilon $ and remains stable in some interval near $T_c$. The jump at $T_c$ and the range of stability of 1RSB solution tend to zero as $p\to2$.

\section{Conclusions \label{Sec:Conc}}

In this paper we investigate the generalized $p$-spin models that contain arbitrary diagonal operators $\hat U$ instead of Ising spins. We focus our attention mainly on the case when $\hat U$ does not have the reflection symmetry (such systems as a whole have no time-reversal symmetry).

We derive general equations that allow to investigate analytically the qualitative behavior of the system near the glass transition at different (continuous) $p$.  The main results are schematically shown in figure ~\ref{fig0}a.

For the quadrupole glass with ${\bf J}=1$ the detailed quantitative analysis is performed.  At $p=3.3$ it is shown that $\mRSB = 1$ at the branch point and the crossover from continuous to discontinuous transition takes place. For $p <2+\Delta p$ (where  $ \Delta p= 0.5$ is a finite value) we get $\lambda_{\rm (\RSB) repl}\leq 0$ for  $T<T_{0}$ (stable 1RSB-solution disappears). This behaviour differs from the corresponding behavior for the conventional $p$-spin Ising glass model.

\section{Acknowledgments}

This work was supported in part by the Russian Foundation for Basic Research (Grant No. 11-02-00341 and 13-02-91177), the Grant of President of Russian Federation for support of Leading Scientific Schools No.~6170.2012.2, NSF Grant DMR 1158666 and Russian Academy of Sciences programs.

\appendix

\section{$W$, $L$, and $B_4$}

 \begin{eqnarray}\label{B11eq:JP}\fl W=\left[\frac{p(p-1)}{2}{q^{(p-2)}_{\RS}}\right]\left\{\langle\hat{U}_{1}^2\hat{U}_{2}^2\rangle-2\langle\hat{U}_{1}^2\hat{U}_{2}\hat{U}_{3}\rangle+
 \langle\hat{U}_{1}\hat{U}_{2}\hat{U}_{3}\hat{U}_{4}\rangle\right\}=
 \\\nonumber
 \int dz^G \left\{\frac{\Tr\left(\hat{U}^2
e^{\hat{\theta}_{\RS}}\right)} {\Tr e^{\hat{\theta}_{\RS}}}-
\left[\frac{\Tr\hat{U} e^{\hat{\theta}_{\RS}}}
{\Tr e^{\hat{\theta}_{\RS}}}\right]^2\right\}^2.
 \end{eqnarray}
Notation used below are built according to the same rules as that in Eq.~(\ref{B11eq:JP}):
 \begin{eqnarray}\label{eqL}
 \fl L={\left[\frac{p(p-1)}{2}{q^{(p-2)}_{\RS}}\right]}^{2}\left\{\langle\hat{U}_{1}^2\hat{U}_{2}\hat{U}_{3}\rangle-
 \langle\hat{U}_{1}\hat{U}_{2}\hat{U}_{3}\hat{U}_{4}\rangle\right\}\geq 0
 \end{eqnarray}
similar to (\ref{B73RS}).

\begin{eqnarray*}\nonumber
\fl t^{6}B_{4}=t^{6}{\left[\frac{p(p-1)}{2}{q^{(p-2)}_{\RS}}\right]}^{3}\left\{\frac{1}{3}\langle\hat{U}_{1}\hat{U}_{2}\hat{U}_{3}\hat{U}_{4}\hat{U}_{5}\hat{U}_{6}\rangle- \langle\hat{U}_{1}^{2}\hat{U}_{2}\hat{U}_{3}\hat{U}_{4}\hat{U}_{5}\rangle+\frac{1}{3}\langle\hat{U}_{1}^{3}\hat{U}_{2}\hat{U}_{3}\hat{U}_{4}\rangle+ \right.
\\
\fl\left.
 \frac{3}{4}\langle\hat{U}_{1}^{2}\hat{U}_{2}^{2}\hat{U}_{3}\hat{U}_{4}\rangle-\frac{1}{2}\langle\hat{U}_{1}^{3}\hat{U}_{2}^{2}\hat{U}_{3}\rangle+
 \frac{1}{12}\langle\hat{U}_{1}^{3}\hat{U}_{2}^{3}\rangle\right\}-
 \frac{t^{2}p(p-1)(p-2)}{12}{q^{(p-3)}_{\RS}}\left[1-t^{2}\frac{3}{2}W\right].
 \end{eqnarray*}

\bibliographystyle{iopart-num}
\bibliography{references}

\end{document}